\pdfoutput=1
\documentclass[aps,prl,reprint,amsmath,amssymb,superscriptaddress]{revtex4-1}
\usepackage{bm}
\usepackage{graphicx}
\usepackage[colorlinks]{hyperref}
\usepackage{mathrsfs}
\usepackage{times}
\usepackage[normalem]{ulem}
\usepackage{soul}

\newcommand{\ket}[1]{|#1\rangle}

\usepackage[normalem]{ulem} 

\newcommand{\beq}{\begin{equation}}
\newcommand{\eeq}{\end{equation}}

\begin{document}

\title{Confined phases of one-dimensional spinless fermions coupled to $Z_2$ gauge theory}

\author{Umberto Borla}
\affiliation{Department of Physics, Technical University of Munich, 85748 Garching, Germany}
\affiliation{Munich Center for Quantum Science and Technology (MCQST), Schellingstr. 4, D-80799 M\"unchen}

\author{Ruben Verresen}
\affiliation{Department of Physics, Technical University of Munich, 85748 Garching, Germany}
\affiliation{Department of Physics, Harvard University, Cambridge, Massachusetts 02138, USA}
\affiliation{Max-Planck-Institute for the Physics of Complex Systems, 01187 Dresden, Germany}

\author{ Fabian Grusdt}
\affiliation{Department of Physics, Technical University of Munich, 85748 Garching, Germany}
\affiliation{Munich Center for Quantum Science and Technology (MCQST), Schellingstr. 4, D-80799 M\"unchen}
\affiliation{Department of Physics and Arnold Sommerfeld Center for Theoretical Physics (ASC), Ludwig-Maximilians-Universit\"at M\"unchen, Theresienstr. 37, M\"unchen D-80333, Germany}

\author{Sergej Moroz}
\affiliation{Department of Physics, Technical University of Munich, 85748 Garching, Germany}
\affiliation{Munich Center for Quantum Science and Technology (MCQST), Schellingstr. 4, D-80799 M\"unchen}

\begin{abstract}
We investigate a quantum many-body lattice system of one-dimensional  spinless fermions interacting with a dynamical $Z_2$ gauge field. The gauge field mediates long-range attraction between fermions resulting in their confinement into bosonic dimers. At strong coupling we develop an exactly solvable effective theory of such dimers with emergent constraints.
Even at generic coupling and fermion density, the model can be rewritten as a local spin chain. Using the Density Matrix Renormalization Group the system is shown to form a Luttinger liquid, indicating the emergence of fractionalized excitations despite the confinement of lattice fermions.
In a finite chain we observe the doubling of the period of Friedel oscillations which paves the way towards experimental detection of confinement in this system. We discuss the possibility of a Mott phase at the commensurate filling $2/3$.
\end{abstract}

\maketitle


\emph{Introduction.}---
Lattice gauge theories, introduced by Wilson in high-energy physics \cite{Wilson1974}, emerge in many condensed matter problems  as low-energy effective theories of exotic quantum phases of matter \cite{wenbook, Sachdevbook, Fradkin2013}. They naturally describe the fractionalization of elementary excitations and deconfined quantum critical points. The simplest lattice gauge theory that has $Z_2$ Ising degrees of freedom was invented by Wegner \cite{Wegner1971}. Being the first example of a system exhibiting topological order, it  provided a paradigm shift in our understanding of phase transitions \cite{Kogut1979}. Its exactly solvable limit, the toric code model \cite{Kitaev2003}, gave a first impetus towards topological quantum computation.

Coupling to gapless dynamical matter can qualitatively change the phase diagram of  a lattice gauge theory. Coupling bosonic (Ising) matter to the Wegner's $Z_2$ gauge theory was undertaken by Fradkin and Shenker already in 1979 \cite{Fradkin1979}, demonstrating that the system exhibits two phases akin to the pure $Z_2$ gauge theory. Models where a $Z_2$ gauge field couples to fermionic matter were studied considerably later. First, motivated by high-$T_c$ cuprate phenomenology, Senhtil and Fisher introduced a two-dimensional $Z_2$ gauge theory coupled to fractionalized fermions and bosons at finite density \cite{Senthil2000}. The fractionalized non-Fermi liquid phase called orthogonal fermions realizes another example of a $Z_2$ gauge theory coupled to fermions and Ising spins \cite{Nandkishore2012}.  More recently, Gazit and collaborators investigated in detail the quantum phase diagram of two-dimensional spinful fermions coupled to the standard $Z_2$ gauge theory using sign-problem-free quantum Monte Carlo simulations \cite{Gazit2017, *Gazit2018, *Gazit2019}. Exactly solvable models of a nonstandard $Z_2$ gauge theory coupled to fermionic matter were constructed \cite{Prosko2019} and argued to exhibit disorder-free localization \cite{Smith2017}. Fermions at finite density coupled to Ising gauge theory without the Gauss law were studied  in \cite{Grover2016, Frank2019}. For related recent work on unconstrained $Z_2$ gauge theories coupled to bosonic matter, see \cite{gonzalez2019intertwined, *PhysRevB.99.045139}. Recently, $Z_2$ gauge fields were also seen to emerge at deconfined quantum critical points in one dimension \cite{PhysRevB.99.075103}.

Studying models with dynamical matter coupled to gauge fields numerically is a challenging task. Recently this has motivated analogue quantum simulations of such problems, using ultracold atoms platforms in particular \cite{Wiese2013, *Zohar2015,*Dalmonte2016,*Mil2019}. Pioneering experimental work in ultracold ions \cite{Martinez2016} has led to the first simulation of string breaking in the Schwinger model 
(QED$_2$) \cite{Schwinger1962}. More recently, a Floquet implementation for $Z_2$ lattice gauge theories coupled to dynamical matter has been proposed \cite{Barbiero2018} and proof-of-principle experiments on a two-component mixture of ultracold atoms have been performed \cite{Schweizer2019, *Gorg2019}. In addition to analogue quantum simulations, advances in quantum computing have stimulated the development of quantum-classical algorithms that are applicable to lattice gauge theories \cite{Klco2018}.

In this Letter we present a study of a one-dimensional quantum $Z_2$ lattice gauge theory coupled to spinless fermions at finite density.
We demonstrate that a local change of basis recasts the model as a spin-1/2 chain without gauge redundancy.
In addition to analytic arguments, we rely on the numerical solution using the Density Matrix Renormalization Group (DMRG) approach in both finite and infinite geometries. At finite coupling the Ising gauge field mediates a linear attractive potential between fermions, confining pairs into bosonic dimer molecules.
As a result, the gauge-invariant fermionic two-point correlation function decays exponentially. On the other hand, our findings suggest that at a generic fermion density, the dimers form a Luttinger liquid. This picture becomes especially tractable in the limit of strong coupling, where dimers are tightly bound hard-core bosonic objects. In this regime, second-order degenerate perturbation theory maps the problem to a constrained model of bosons with short-range repulsive interactions which was solved analytically via the Bethe ansatz \cite{Alcaraz1999}. Recent experiments with one-dimensional chains of Rydberg atoms \cite{Bernien2017} reignited theoretical work on lattice models with extended hard-core constraints \cite{Turner2018, *Chepiga2019, *Giudici2019, *Verresen2019, *surace2019lattice}. Our work demonstrates that such constraints emerge naturally at low energies in discrete lattice gauge theories with confined fermions.
 
\begin{figure}[ht]
	\includegraphics[width=0.85\linewidth]{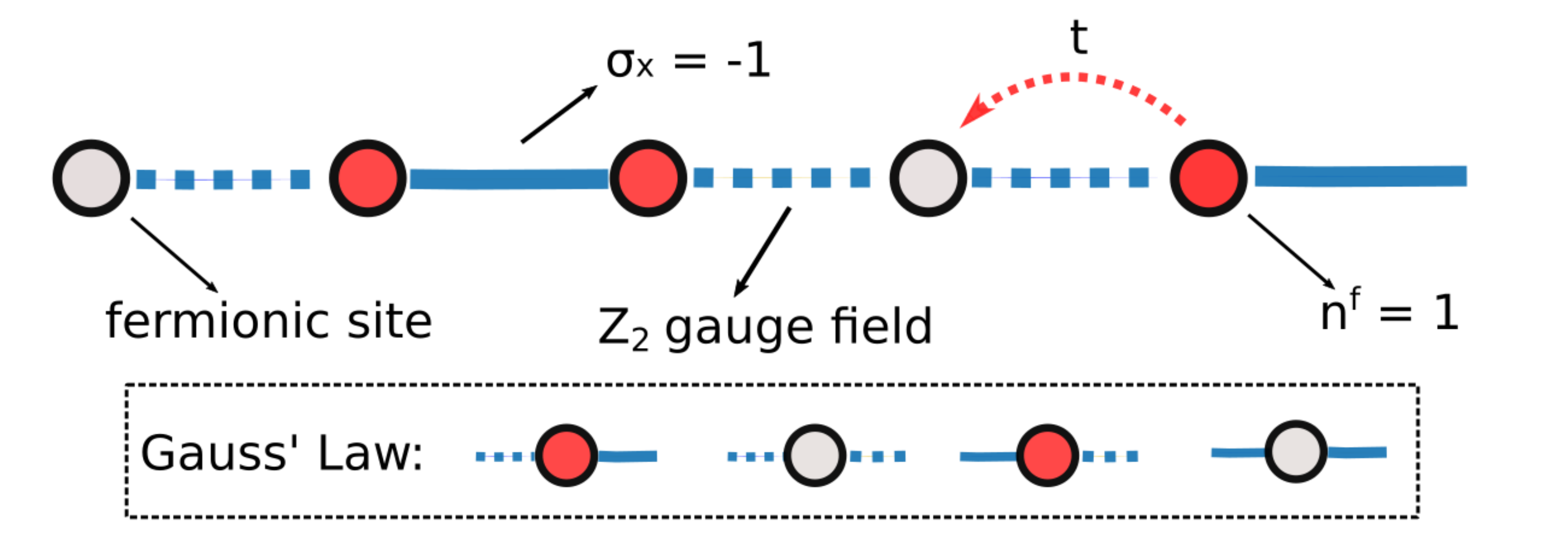}
	\caption{Fermions (red) that occupy sites interact with the $Z_2$ gauge field (blue) defined on links. Any physical configuration satisfies the Gauss law.}
	\label{fig:model}
\end{figure}

\emph{The model.}--- We consider a chain where spinless fermions $c_i$ live on sites and $Z_2$ Ising gauge fields are defined on links, see Fig. \ref{fig:model}. The quantum Hamiltonian of the system is
\beq \label{H}
H=-t\sum_{i}(c_i^{\dagger}\, \sigma^{z \vphantom \dagger}_{i,\,i+1}\, c_{i+1}^{\vphantom \dagger}+\text{h.c.}) -h\sum_{i}\sigma^x_{i,\,i+1}.
\eeq
Fermions are coupled minimally to the gauge field via the Ising version of the Peierls substitution. The second term in Eq. \eqref{H}---the discrete version of the electric term in electrodynamics---induces transitions between the gauge Ising spins. A related lattice problem with gapless bosonic matter was investigated in \cite{PhysRevB.84.235148}.

The Hamiltonian commutes with local $Z_2$ generators $G_i = \sigma^x_{i-1,\,i}\, (-1)^{n^f_i} \sigma^x_{i,\,i+1}$, where $n^f_i = c^\dagger_i c_i$ is the number operator. As a consequence, each choice $G_i = \pm 1$ corresponds to an independent sector of the Hilbert space. We choose to minimize the Hamiltonian \eqref{H} under the set of constraints $G_i=1$. This leads to an Ising version of the Gauss law. Importantly, in a closed chain the gauge constraint implies that the fermion parity of any state in the physical Hilbert space is even, removing all fermionic excitations from the energy spectrum. In a finite chain, the allowed fermion parity depends on whether the chain ends with a site or a link \cite{supmat}.

We briefly contrast our model to the paradigmatic Schwinger model \cite{Schwinger1962}---a one-dimensional $U(1)$ gauge theory exhibiting confinement and string-breaking \cite{Coleman75,*Rothe79,*Manton85}. Considerable work \cite{Kogut75,*Hamer82,*Schiller83,*Sriganesh00,*Byrnes02,*Banuls13,*Buyens14,*Rico14,*Dalmonte16,*Ercolessi2018,*Sala18,*Magnifico19,*Magnifico19b} has been done on its lattice discretization as well as its quantum link versions where the gauge field is finite-dimensional. 
In fact, the phenomenology of the Schwinger model with a theta-angle $\theta = \pi$ resembles that of a quantum link model where the
electric field only has two levels \cite{Banerjee12,*Surace19}. Intriguingly, that effective model is superficially similar to Eq.~\eqref{H}, with three important differences: (1) $\sigma^z_{i,i+1}$ in the hopping is replaced by a raising operator $\sigma^+_{i,i+1}$, (2) the Gauss law is not translation-invariant, and (3) the Gauss law only allows
three \footnote{The electric fields can only \emph{exit}---not \emph{enter}---a charge.} of the four states in Fig.~\ref{fig:model}. Consequently, the physics is fundamentally different, leading to, e.g., the confinement of fermions into charge-zero objects rather than charge-two dimers.

The energy spectrum of Eq.~\eqref{H} is symmetric under the transformation $t\to -t$ since one can perform a unitary rotation acting on the links as $\sigma_x \to \sigma_x$, $\sigma_y \to -\sigma_y$, $\sigma_z \to -\sigma_z$, which flips the sign of $t$ in the Hamiltonian but preserves the Gauss law.
A similar argument implies that in a periodic chain the spectrum is invariant under $h\to -h$. 
In the following we will therefore only consider $t, h \ge 0$.

In addition to local $Z_2$ gauge invariance, the model exhibits a global $U(1)$ symmetry acting only on fermions $c_i\to e^{i\alpha}c_i$. We can thus work in the grand canonical ensemble with the Hamiltonian $H-\mu \sum_i n_i^f$ and change the fermion density by tuning the chemical potential $\mu$.
 
At $h=0$, the gauge field decouples and the model reduces to free fermions. Formally, this can be demonstrated by introducing $Z_2$ gauge-invariant but non-local fermion operators $f_i= c_i \prod_{j\ge i} \sigma^z_{j,j+1}$. In terms of $f_i$, the Hamiltonian \eqref{H} at $h=0$ reduces to the canonical model of non-interacting fermions, whose ground state at finite filling is a free Fermi gas. However, even in this case, due to the $Z_2$ Gauss law, a single $f_i^\dagger$ does not create a physical excitation in a closed chain. We say that the model has emergent fermions $f_i$, which can only be created in pairs. 

\emph{Confinement.}---
The gauge constraint ensures that bare $c_i$ fermions are connected by electric strings with $\sigma_x = -1$. Since for $h>0$ the electric term introduces an energy cost
for such lines, the bare fermions are expected to become confined into dimers. Later, we confirm this by showing that for any $h \neq 0$, the two-point correlator $\langle f_i^{\dagger} f_{i+d}^{\vphantom \dagger}\rangle$ decays exponentially fast with $d$.
In the limit $h \to \infty$, we will derive an effective integrable Hamiltonian for the bound states, which forms a Luttinger liquid.

Remarkably, 
the low-energy theory of Eq.~\eqref{H} is a Luttinger liquid for a generic value of $h$, with a smoothly varying Luttinger liquid parameter. At first sight, this might seem in contradiction with the observation that the bare $f_i$ fermions are confined and have exponentially decaying correlators when $h\neq 0$. However, in the universal low-energy regime, there will be new emergent deconfined fermions \footnote{As the Luttinger liquid parameter drifts from the free-fermion value, the emergent fermions are not quasiparticles, obfuscating their meaning.}.

To see this, it is useful to consider the bosonization of the electric term of the Hamiltonian \eqref{H}. Due to the Gauss law, at any site $i$ we have
$
\sigma^x_{i-1, i} = (-1)^{n^f_i} \sigma^x_{i, i+1} 
$. 
Applying this relation successively at each site on the right of site $i$, and assuming that $\sigma^x$ at infinity (or at the boundary) is $+1$, we find that
\begin{equation} \label{xJW}
\sigma^x_{i-1, i} = (-1)^{\sum_{j\ge i}n^f_j} = e^{i\pi\sum_{j \ge i}n^f_j} \approx e^{i\pi\int_{x>x_i} \rho_F(x)},
\end{equation}
taking the continuum limit in the last step. Hence, $\sigma^x$ can be replaced with a non-local operator that only depends on the density of fermions. Using bosonization $\rho_F(x) \rightarrow \rho^0_{F} -\partial_x \phi(x)/\pi$ \footnote{Here we follow conventions of \cite{Giamarchibook, Sachdevbook}, where single fermion corresponds to a $\pi$-kink of the field $\phi$.}
, the electric term  of the Hamiltonian \eqref{H} becomes
\begin{equation} \label{bos}
-h \int dx e^{i\pi\int_{y>x} dy \rho_F(y) }   \rightarrow -h \int dx  \cos{(k_F x-\phi(x))},
\end{equation}
with Fermi momentum $k_F=\pi \rho^0_{F}$. The cosine might seem to energetically punish $\pi$ kinks---which are exactly the fermionic excitations---and favors $2\pi$ kinks---giving rise to dimers as the effective degree of freedom.  However, the spatial dependence of Eq.~\eqref{bos} implies that the perturbation is not RG-relevant: near the free-fermion point, the $U(1)$-symmetric relevant perturbations are $\cos(\phi)$ and $\cos(2\phi)$, with momentum $k=k_F$ and $k=2k_F$, respectively. Due to the lattice translation symmetry of the Hamiltonian, neither of these terms can be generated at any finite filling fraction (i.e., $0 < k_F < \pi$) \cite{Affleck:1988mua}. As a result, the system remains a Luttinger liquid as we tune $h$, with a flowing Luttinger parameter due to the symmetry-allowed marginal perturbation $(\partial \phi)^2$.
This agrees with the analysis of Ref.~\cite{PhysRevB.84.235148} of $Z_2$ instanton effects in the path-integral formalism.

Any Luttinger liquid has emergent deconfined fermionic excitations $e^{i(\phi\pm\theta)}$. The special property of $h=0$ is that these emergent excitations coincide with the \emph{lattice} fermions $f_i$. When $h\neq 0$, the lattice-continuum correspondence between $f_i$ and the low-energy field operators $\phi,\theta$ is modified \cite{supmat}. While the bare fermions $f_i$ are confined into dimers, we will argue below that the interactions between the latter lead to the formation of a collective Luttinger liquid phase whose elementary excitations are fractionalized. In fact, we now show that the model is equivalent to a spin-$1/2$ chain for any value of $h$, illustrating that even for $h=0$, the fermionic excitations can be thought of as collective deconfined fermionic spinons---in this limiting case created by $f_i$.

To rewrite the model \eqref{H} as a local spin-$1/2$ chain, we introduce Majorana operators $\gamma_i=c_i^\dagger+c_i^{\vphantom \dagger}$ and $\tilde \gamma_i= i (c_i^\dagger- c_i^{\vphantom \dagger})$. After some algebra \cite{supmat}, the Hamiltonian \eqref{H} is transformed to
\beq \label{KW}
H=-\frac{t}{2}\sum_{i}\left(1 - X_{i-1,i} X_{i+1, i+2}\right)Z_{i,i+1} -h\sum_i X_{i,i+1},
\eeq
where $X_{i,i+1}=\sigma^x_{i,i+1}$ and $ Z_{i,i+1}=-i \tilde \gamma_i \sigma^z_{i,i+1}\gamma_{i+1}$ are gauge-invariant local operators \cite{Radicevic2018}.
Since $n^f_i=(1- X_{i-1,i}  X_{i, i+1} )/2$, confinement of fermions manifests itself in this formulation as confinement of domain walls, whose hopping is governed by the first term in Eq. \eqref{KW} \cite{suzuki1971relationship, *Keating2004, *PhysRevLett.93.056402, *PhysRevLett.103.020506, *zeng2019quantum,  *PhysRevA.99.060101}. The gauge-invariant Majorana fermions correspond to such operators as $ X_n  Z_{n+1}  Z_{n+2} \cdots$. We emphasize that Eq.~\eqref{KW} is obtained by a local change of variables without any gauge-fixing; this is consistent with the field-theoretic perspective that a $Z_2$ gauged Dirac fermion is a compact boson \cite{Karch:2019lnn}.

Using the TeNPy Library \cite{tenpy}, DMRG simulations presented in this Letter were performed for either the original constrained model \eqref{H} or the unconstrained model \eqref{KW}. 
\begin{figure}[ht]
	\includegraphics[width=0.85\linewidth]{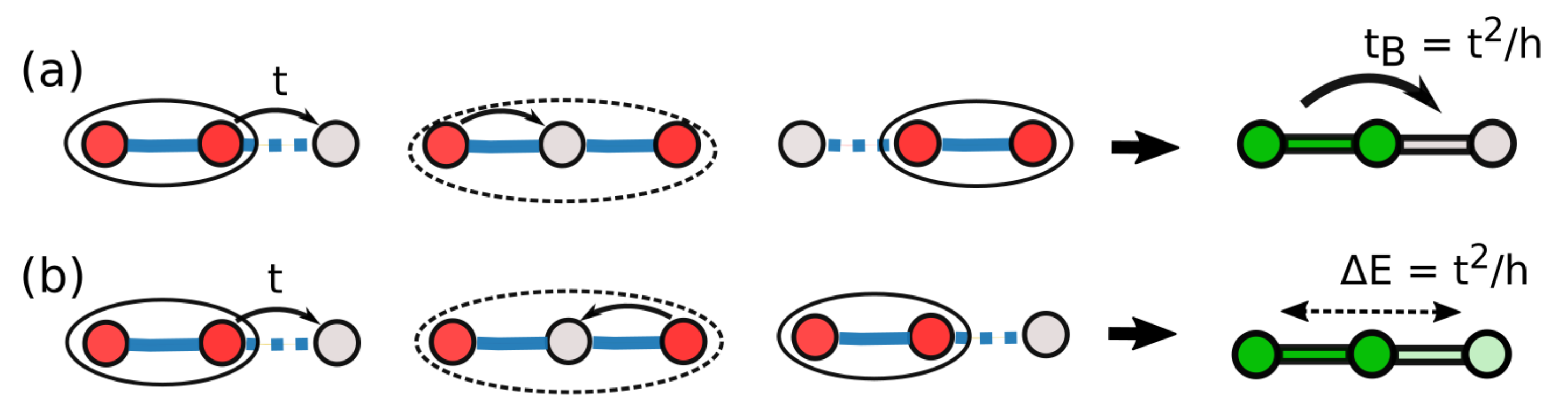}
	\caption{Effective dimer model from the second order perturbation theory: (a)  effective hopping of a dimer,  (b) dimer length fluctuation that decreases the energy of a dimer.}
	\label{fig:processes}
\end{figure}
\begin{figure*}
	\includegraphics[width=\linewidth]{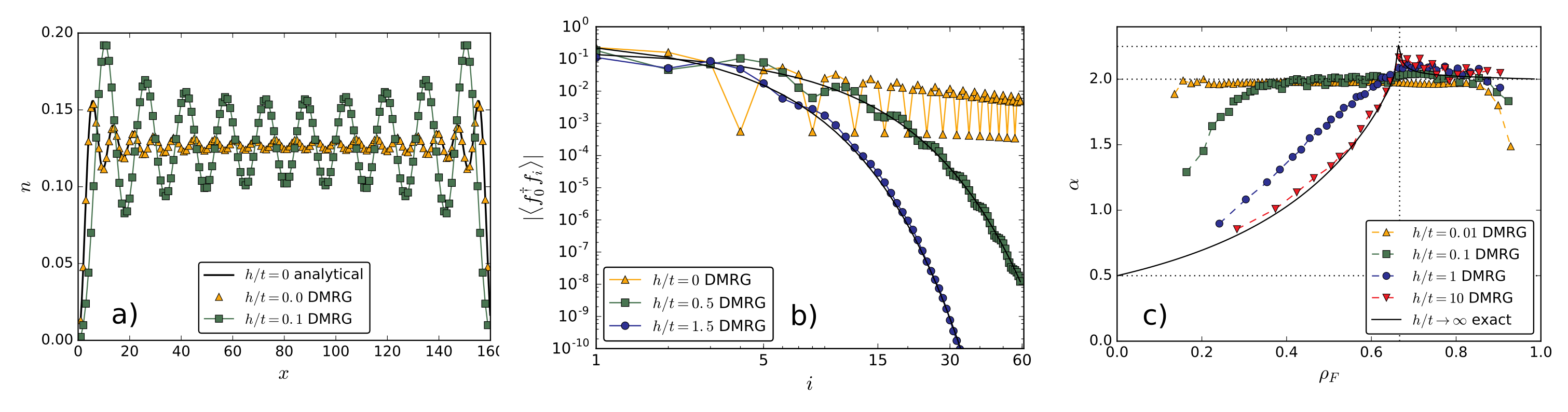}
	\caption{(a) Friedel oscillations in a chain of length $L=160$ at filling $\rho_F = 1/8$. At $h=0$ the oscillations have the wave-vector $2k_F$, but for $h>0$ the wave-vector is reduced to $k_F$ resulting in the doubling of the period. (b) Absolute value of the fermion correlator $\langle f_0^\dagger f_i \rangle$  for different values of the gauge coupling $h$. At $h=0$ the correlators decay algebraically. For $h>0$ the decay is exponential, indicating confinement of fermions $f_i$. The black lines are exponentials, obtained by fitting the data. (c) Exponent of the power-law decay of the pair-pair correlator  $\langle b_i^\dagger b_j \rangle$. The numerical results at $h=10$ are in a good agreement with the predictions from the exactly-solvable bosonic model \eqref{EH}. For figures (a) and (b) the ground state is obtained by applying finite DMRG to the original Hamiltonian \eqref{H}. For figure (c), iDMRG on the dual spin model \eqref{KW} was used.} 
	\label{fig:combined}
\end{figure*}

\emph{Effective theory of compact dimers at $ h\gg t$.}---
When the electric coupling $h$ is much larger than the hopping $t$,
the fundamental low-energy degrees of freedom are tightly bound pairs of
fermions. Such bosonic molecules live on the dual lattice formed by the links $i^*=\left( i, i+1\right)$ of the original lattice, and are created by the gauge-invariant operator $b^{\dagger}_{i^*} = c^{\dagger}_i \sigma^z_{i^*} c^{\dagger}_{i+1}$. Due to Pauli's principle, the dimers are hardcore bosons which are not allowed to occupy neighboring sites: the effective theory has a constrained Hilbert space. Here we write the corresponding low-energy model due to second order degenerate perturbation theory in the hopping parameter $t$. First, a hopping of a dimer between neighboring sites of strength $t_B = t^2/2h$ is generated by the process illustrated in Fig. \ref{fig:processes} (a). Second, a pair can reduce its energy if one of the fermions hops once, and then hops back to the initial state (Fig. \ref{fig:processes} (b)). Since such a process is inhibited when two pairs sit next to each other, a next-nearest-neighbor repulsion of dimers is induced. The strength of this repulsion is equal to $U_B=2 t_B$. The effective Hamiltonian is
\begin{equation} \label{EH}
H_B =  \sum_{j^*} \mathcal{P}_1 \left[ -t_B \left( b^{\dagger}_{j^*} b_{j^*+1}^{\vphantom \dagger} + h.c.\right) + U_B \,n^B_{j^*} n^B_{j^*+2}\right] \mathcal{P}_1,  
\end{equation}
where $n^B_{j^*}=b^{\dagger}_{j^*} b^{\vphantom \dagger}_{j^*}$ and $\mathcal{P}_1$ is a projector that enforces the hard-core constraints $b^{\dagger 2}_{j^*}=b^\dagger_{j^*}b^\dagger_{j^*+1}=0$.

The model \eqref{EH} can be mapped onto a constrained XXZ spin chain \cite{supmat} which was solved analytically using the Bethe ansatz \cite{Alcaraz1999}. From that solution one finds that at a generic filling that $\langle b^\dagger_{i^{*}} b_{j^*}\rangle$ decays algebraically $\sim|i^*- j^*|^{-\alpha}$ with a density-dependent exponent
\begin{equation}
\alpha = \frac{1}{2(1-\rho_B)^2\,\eta_{\rho_B}^2},
\label{alpha_alcaraz}
\end{equation}
fixing the Luttinger parameter $K=1/(2\alpha)$ for the effective model \eqref{EH}.
Here $\rho_B = \rho_F /2 \le 1/2$ is the boson density and the density-dependent parameter $\eta_{\rho_B}$ is determined by solving a certain system of integral equations \cite{Alcaraz1999, Karnaukhov2002}, see \cite{supmat} for a summary. At low densities, $\eta_{\rho_B}\to 1$ such that $\alpha = 1/2$ as expected for non-interacting hardcore-bosons ($K=1$). As the density increases towards $\rho_B=1/3$, the particles start to feel more repulsive interactions, and $\alpha$ increases monotonically. Due to the constraint, the density $\rho_B = 1/3$ is special and can be considered ``half filling" since then there is one boson on every other bond allowed by the constraint. At this filling, for weak repulsion $U_B/t_B<2$, the model \eqref{EH} is a Luttinger liquid, but as the repulsion parameter is increased to $U_B/t_B>2$ it forms a $Z_3$ Mott insulator \cite{supmat}. Hence, at leading order in the $t/h$ expansion, this model lies on the interface between these two phases. Note that at $\rho_B=1/2$, the constraints force the model \eqref{EH} to be a $Z_2$ insulator.

As explained in detail in \cite{supmat}, the effective model for the dimers \eqref{EH} can be related to the $SU(2)$ invariant spin-$1/2$ Heisenberg chain in squeezed space whose local excitations are known to fractionalize into deconfined fermionic spinons \cite{Giamarchibook, Fradkin2013}.


\emph{Friedel oscillations.}---
A clear indication that the effective degrees of freedom of our system are bosonic dimers follows from the Friedel oscillations observed in a finite chain with open boundary conditions. For free fermions ($h=0$), we find oscillations of the density with the inverse period equal to $2k_F$ \cite{friedel1958metallic}. In the limit $ h \gg t$, one expects a doubling of the period of Friedel oscillations because the density of hardcore dimers is half the density of fermions. Our DMRG results suggest that the doubling occurs for any $h\ne 0$ (see Fig. \hyperref[fig:combined]{3a}). Cold atom experiments can directly measure the periodicity of Friedel oscillations and thus detect signatures of confined fermions in this model.


\emph{Correlation functions.}---
Due to Elitzur's theorem, only gauge-invariant observables can have non-zero expectation values. In our model, we construct such quantities by working with the gauge-invariant fermions $f_i$. 

From our DMRG simulation we first extract the equal-time fermionic correlator $\langle f_i^\dagger f_j^{\vphantom \dagger} \rangle$. As illustrated in Fig. \hyperref[fig:combined]{3b}, the power-law behavior at $h=0$ changes to an exponential decay for $h>0$. We attribute this behavior to the long-range confining interaction between $Z_2$ charges.

Next, we compute the dimer-dimer equal-time correlation function $\langle b_i^\dagger b_j^{\vphantom \dagger} \rangle$. Our DMRG results indicate that at a generic gauge coupling $h$ and filling $\rho_F$, the correlation function falls off algebraically with exponents shown in Fig. \hyperref[fig:combined]{3c}. In the regime $ h \gg t$, our findings are in good agreement with Eq. \eqref{alpha_alcaraz} derived for the effective model \eqref{EH}.
For lower values of $h$, the curves $\alpha(\rho_F)$ deviate from the predictions of the effective model since the dimers no longer have a fixed unit length. 
Qualitatively, compared to the regime $ h \gg t$, we expect $\alpha$ to increase more rapidly as a function of $\rho_F$ since for larger dimers the mutual repulsive interactions kick in at lower densities. We also observe that in the regime $t\gg h$ the exponent of the dimer-dimer correlator saturates quickly to $\alpha = 2$, the value for the free-fermion case ($h=0$) which is perturbatively stable due to Eq. \eqref{bos} being a marginal perturbation. In the light of these observations, it is tempting to think of our system as a Luttinger liquid of dimers even away from the limit $ h \gg t$, where compact dimers are proper low-energy degrees of freedom. Our two-body calculation of the average dimer size indicate that it is of order unity  even for relatively small values of $h/t$ \cite{supmat}. In addition, in all studied cases the entanglement entropy scaling extracted from DMRG resulted in the central charge $c=1$ as expected for a Luttinger liquid \cite{Calabrese2004, *Pollmann2009}.  


\emph{Outlook}---Does the Mott state appear at $2/3$ filling? Second order perturbation theory fixed the ratio $U_B/t_B=2$ in the Hamiltonian \eqref{EH}, which at the filling $\rho_B=1/3$ places the effective model exactly at the transition point between the Luttinger liquid and the $Z_3$ Mott state. Higher-order perturbation theory can modify this ratio and generate further-range hoppings and interactions. Such a calculation would shed light on the fate of the filling $\rho_F=2/3$ in the regime $h\gg t$. For $h\sim t$, it would be interesting to numerically investigate the nature of the ground state at this commensurate filling.

It has been realized recently that confining theories can exhibit non-ergodic behavior \cite{Kormos2016, *Brenes2018, *James2019, *Robinson2019, *pai2019}. The model investigated here might provide a new means to explore quantum scarred states \cite{Turner2018Nat, *Turner2018PRB, *Moudgalya2018, *PhysRevB.98.235156,  *Lin2019, *Schecter2019}.

The model studied in this Letter can be realized experimentally using ultracold atoms in optical lattices. One possibility to implement the coupling of fermions to a $Z_2$ lattice gauge field is to use the Floquet scheme proposed and demonstrated in Refs. \cite{Barbiero2018,  Schweizer2019}. An alternative is to use doped quantum magnets: consider a 1D Fermi-Hubbard model at strong coupling $U \gg t$ in the presence of a staggered Zeeman field $h \sum_j (-1)^j \hat{S}^z_j$ with $h \gg J$ exceeding the super-exchange energy $J =4 t^2 / U$. This model can be implemented by confining two pseudospin states with different magnetic moments (e.g. 40K \cite{cheuk2016observation}) to a zig-zag optical lattice and adding a perpendicular magnetic gradient. At half-filling, the ground state is a classical N\'eel state. By adiabatically decreasing the density, mobile holes can be introduced and the ground state can be adiabatically prepared \cite{hilker2017revealing}. Because the strong Zeeman field $h \gg J$ suppresses fluctuations of the spins in the $x-y$ plane, pairs of holes are connected by a string of misaligned spins which can be mapped onto the $Z_2$ electric field lines considered in our model. In the limit $U \to \infty$, this model is equivalent to Eq. \eqref{H}, with holes corresponding to the fermionic $Z_2$ charges. 

\emph{Note added}---After our preprint appeared, Ref \cite{Iadecola2019} constructed two towers of exact many-body scar states in the spin $1/2$ model \eqref{KW}. Moreover, initial states producing periodic revivals were identified.


%

\begin{acknowledgments}
\emph{Acknowledgements.}---We acknowledge useful discussions with Ian Affleck, Monika Aidelsburger, Luca Barbiero, Annabelle Bohrdt, Giuliano Giudici, Johannes Hauschild, Lesik Motrunich, Frank Pollmann, Kirill Shtengel, Senthil Todadri, Carl Turner,  Julien Vidal and participants of Nordita program ``Effective Theories of Quantum Phases of Matter''.  Our work ~is funded by the Deutsche Forschungsgemeinschaft (DFG, German Research Foundation) under Emmy Noether Programme grant no.~MO 3013/1-1 and under Germany's Excellence Strategy - EXC-2111 - 390814868. F.G. acknowledges support from the Technical University of Munich - Institute for Advanced Study, funded by the German Excellence Initiative and the European Union FP7 under grant agreement 291763, from the DFG grant No. KN 1254/1-1, and DFG TRR80 (Project F8).  R.V. was partly supported by the German Research Foundation (DFG) through the Collaborative Research Center SFB 1143. RV acknowledges support from the Harvard Quantum Initiative Postdoctoral Fellowship in Science and Engineering and from a Simons Foundation grant (\#376207, Ashvin Vishwanath).
\end{acknowledgments}

\bibliography{library}

\clearpage
\begin{widetext}
\setcounter{page}{1}
\renewcommand{\theequation}{S\arabic{equation}}
\renewcommand{\thefigure}{S\arabic{figure}}
\setcounter{equation}{0}
\setcounter{figure}{0}

\section{Supplemental material: Confined phases of one-dimensional spinless fermions coupled to $Z_2$ gauge theory}
\subsection{Fermion parity in a finite chain}

In a closed chain the Gauss law enforces that the physical Hilbert space contains only states with even number of fermions, i.e., the fermion parity $P=(-1)^{\sum_{i}n_i^f}=1$. In a finite chain the allowed fermion parity depends on whether the chain ends with a site or a link. In the case of a link-boundary, an odd number of fermions is not prohibited by the Gauss law (see Fig. \ref{fig:boundaries} a) and the fermion parity can be either odd or even. On the other hand, for the site-boundary the Gauss law constraint must be modified at the edge. If one imposes at the left boundary  $G_1=(-1)^{n^f_1} \sigma^x_{1,2}=1$ and right boundary $G_L=(-1)^{n^f_L} \sigma^x_{L-1,L}=1$, the fermions parity must be even (see Fig. \ref{fig:boundaries} b).

\begin{figure}[ht]
	\includegraphics[width=0.5\linewidth]{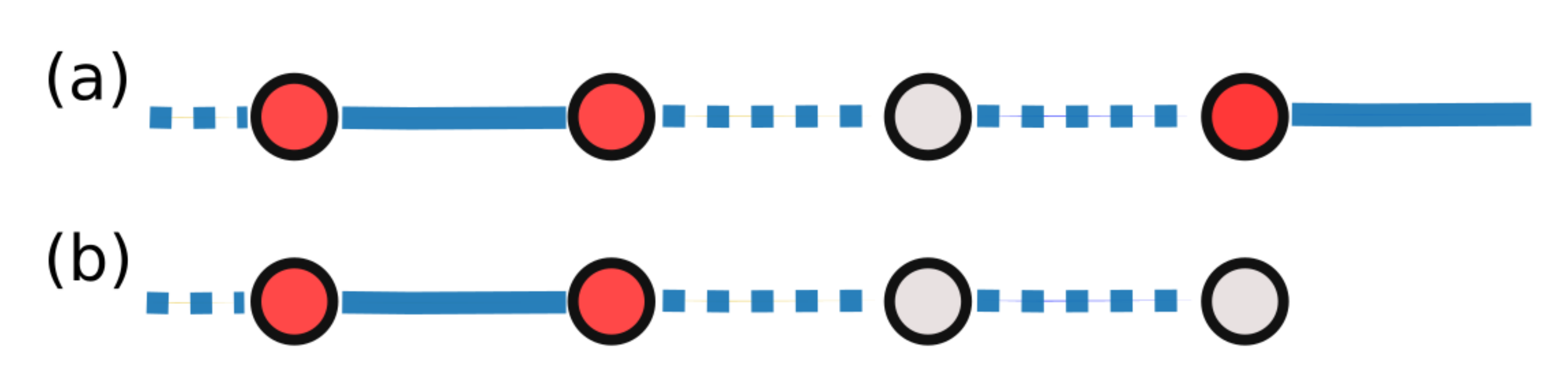}
	\caption{Chain with link-boundary (a) and site-boundary (b) on the right. In the first case, an unpaired fermion can be accommodated, since it is possible to connect it to the boundary with an electric string that terminates on the last link.}
	\label{fig:boundaries}
\end{figure}

\subsection{Lattice-continuum correspondence}

For the free-fermion case, there is a well-known correspondence between the lattice fermion operator and the continuum fields. This correspondence continues to hold in the well-explored scenario where local quartic interactions are introduced \cite{Giamarchibook}. However, if the fermion is coupled to a gauge field, there is a qualitatively new perturbation possible, considered in the model \eqref{H} of the main text. We claim that if $h \neq 0$, the lattice-continuum correspondence changes, such that the lattice fermion $c_i$ (or better, its gauge-invariant non-local version $f_i$) no longer corresponds to $e^{i(\varphi \pm \theta)}$. Indeed, this is already evidenced by the fact that the former lattice operator is numerically observed to have exponentially decaying correlations, see Fig. 3 b), whereas $\langle e^{-i(\varphi(x) + \theta(x)} e^{i(\varphi(0) + \theta(0)} \rangle$ is algebraically decaying within the Luttinger liquid framework.

To better understand the failure of the usual correspondence, it is instructive to consider the equivalent bosonic formulation of the model as shown in Eq. \eqref{KW}. As mentioned in the main text, the lattice fermionic operator is given by $X_n Z_{n+1} Z_{n+2} \cdots$, corresponding to the aforementioned continuum field if $h=0$. Equivalently, the domain wall operator $Z_n Z_{n+1} Z_{n+2} \cdots$ corresponds to $e^{i\varphi(x)}$ (again, if $h=0$); we now clarify why this correspondence breaks down for $h \neq 0$. The reason that they should correspond for $h=0$ is dictated by very general symmetry properties: the lattice $Z_2$ symmetry is generated by the global operator $\prod_n Z_n$, whereas the continuum $Z_2$ symmetry $\theta \to - \theta$ is generated by $e^{i \int \partial_x \varphi(x) \mathrm dx}$ (this is a simple consequence of the fact that $\theta(x)$ and $\partial_x \varphi(x)$ are conjugate fields). Such symmetry matching is a general and powerful method for creating lattice-continuum correspondences. This also explains why things change for $h\neq 0$: the lattice $Z_2$ symmetry is explicitly broken by the last term in Eq. \eqref{KW}; instead, the system develops an emergent $Z_2$ symmetry at low energies (due to the perturbation not being relevant, as discussed in the main text). There is thus no more reason that the lattice domain wall operator should correspond to the continuum field $e^{i\varphi(x)}$.

\subsection{From $Z_2$ gauge theory with fermions to a local spin $1/2$ model}
The Hamiltonian \eqref{H} is written in terms of fermionic degrees of freedom that are not $Z_2$ gauge invariant.
Here we demonstrate how to cast the Hamiltonian into a local form in terms of gauge-invariant observables.

First, it is convenient to introduce Majorana variables $\gamma_i=c_i^\dagger+ c_i$ and $\tilde \gamma_i = i (c_i^\dagger-c_i)$, or equivalently, $c_i=(\gamma_i+i \tilde \gamma_i)/2$  and $c_i^\dagger=(\gamma_i-i \tilde \gamma_i)/2$. In terms of these $\Pi_i=(-1)^{n_i^f}=i \tilde \gamma_i \gamma_i$. Now, using the gauge condition $G_i = +1$, we can rewrite the kinetic term as

\beq
\begin{aligned}
-\left(c_{i}^{\dagger} \sigma^z_{i,i+1} c_{i+1}^{\vphantom \dagger}+\text{h. c.}\right) &=\frac{1}{2}\left(i \tilde{\gamma}_{i} \sigma^z_{i,i+1} \gamma_{i+1}+i \tilde{\gamma}_{i+1}  \gamma_{i} \sigma^z_{i,i+1} G_{i} G_{i+1}\right) \\ 
&=\frac{1}{2}\left(i \tilde{\gamma}_{i} \sigma^z_{i,i+1} \gamma_{i+1}+i \underbrace{\tilde{\gamma}_{i+1} \gamma_{i} \Pi_{i} \Pi_{i+1}}_{=-\tilde{\gamma}_{i} \gamma_{i+1}} \sigma^x_{i-1, i} \sigma^z_{i,i+1} \sigma^x_{i+1, i+2}\right) \\ 
&=\frac{1}{2} i \tilde{\gamma}_{i} \gamma_{i+1}\left(\sigma^z_{i,i+1}-\sigma^x_{i-1, i} \sigma^z_{i,i+1} \sigma^x_{i+1, i+2}\right). 
\end{aligned}
\eeq
Thus far, we have managed to rewrite the Hamiltonian \eqref{H} as
\beq
H=\sum_{i}\frac{t}{2}\left(\sigma^z_{i, i+1}-\sigma^x_{i-1, i} \sigma^z_{i, i+1} \sigma^x_{i+1, i+2}\right) i \tilde{\gamma}_{i} \gamma_{i+1}-h\sum_i \sigma^x_{i, i+1}.
\eeq
In terms of local gauge-invariant operators $X_{i,i+1}=\sigma^x_{i,i+1}$ and $ Z_{i,i+1}=-i \tilde \gamma_i \sigma^z_{i,i+1}\gamma_{i+1}$ the Hamiltonian can be written as
\beq \label{RH}
H=-\frac{t}{2}\sum_{i}\left(Z_{i,i+1} - X_{i-1,i} Z_{i,i+1} X_{i+1, i+2}\right)-h\sum_i X_{i,i+1}.
\eeq
Using the Gauss law, the chemical potential term $-\mu \sum_i n^f_i$ becomes bilocal in this formulation $-\mu \sum_i (1- X_{i-1,i}  X_{i, i+1} )/2$.

\subsection{From constrained XXZ chain to constrained bosons} \label{SMA}
We start from the Hamiltonian of the constrained XXZ chain analyzed in \cite{Alcaraz1999}
\beq \label{HXXZ}
H_{XXZ}=-\frac{1}{2} \sum_{i} P_{l}\left(\sigma_{i}^{x} \sigma_{i+1}^{x}+\sigma_{i}^{y} \sigma_{i+1}^{y}+\Delta \sigma_{i}^{z} \sigma_{i+l+1}^{z}- h \sigma^z_i \right) P_{l},
\eeq
where $P_l$ projects out the states, where two spins up are at a distance that is smaller or equal to $l$ lattice spacings. We introduce hard-core boson operators
\beq
b_i^\dagger=\frac 1 2 \big(\sigma_i^x+i \sigma_i^y \big), \qquad b_i=\frac 1 2 \big(\sigma_i^x-i \sigma_i^y \big)
\eeq
which implies $n^B_i= b_i^\dagger b_i=(1+\sigma_i ^z)/2$. Up to a constant shift,  in terms of these operators the Hamiltonian \eqref{HXXZ} reads
\beq
H=- \sum_{i} \mathcal{P}_{l}\left[ (b_i^\dagger b_{i+1}^{\vphantom \dagger}+h.c.) + 2 \Delta n^B_i n^B_{i+l+1}+ 2(h-\Delta) n^B_i \right] \mathcal{P}_{l},
\eeq
where the projector $\mathcal{P}_l$ forbids two bosons at distances smaller or equal to $l$. For $l=1$ the resulting constrained bosonic model has a unit nearest-neighbor hopping $t_B=1$,  the next-nearest-neighbor interaction of strength $U_B/t_B=-2\Delta$ and the chemical potential $\mu_B/t_B=2(h-\Delta)$.

\subsection{How to determine $\eta_{\rho_B}$ in Eq. \eqref{alpha_alcaraz}}
As demonstrated in the previous subsection, the constrained bosonic model \eqref{EH} introduced in the main text can be mapped onto the constrained XXZ chain \eqref{HXXZ} which was solved using the Bethe ansatz \cite{Alcaraz1999}.
Employing this mapping, for $U_B=0$  (that corresponds to $\Delta = 0$ in the constrained XXZ chain) the boson-boson correlator decays as $\langle b^\dagger_{i^{*}} b_{j^*}\rangle \sim |i^{*}-j^{*}|^{-\alpha}$ with $\alpha = (1-\rho_B)^{-2}/2$.
At $U_B\ne 0$ ($\Delta\neq 0$) the exponent is
\begin{equation} \label{alphaee}
\alpha =\frac 1 2 \,(1-\rho_B)^{-2} \eta_{\rho_B}^{-2},
\end{equation}
where the density-dependent parameter $\eta_{\rho_B}$ can be obtained by solving a system of integral equations derived in \cite{Alcaraz1999, Karnaukhov2002}.
In particular, for $U_B>2t_B$ ($ \Delta <-1$) one can use the parametrization $\Delta = -\cosh(\lambda)$ and the equations to solve are 
\begin{align}
1 &= \eta(U) + \frac{1}{2\pi} \int_{-U_0}^{U_0}\frac{\sinh(2\lambda) \eta(U')}{\cosh(2\lambda)-\cos(U-U')}dU', \label{etaeq} \\
Q(U) & = \frac{1}{2\pi} \frac{\sinh \lambda}{\cosh\lambda-\cos U}-\frac{1}{2\pi} \int_{-U_0}^{U_0} \frac{\sinh(2 \lambda) Q(U')}{\cosh(2 \lambda)-\cos(U-U')}dU',  \label{etaeqa} \\
\int_{-U_0}^{U_0} Q(U) dU &=\left\{
                \begin{array}{ll}
                  \frac{\rho_B}{1-\rho_B}, \qquad 0\le \rho_B \le \frac 1 3, \\
                  \frac{1-2\rho_B}{1-\rho_B}, \qquad \frac 1 3 \le \rho_B \le \frac 1 2,
                \end{array}
              \right. \label{etaeqb}
\end{align} 
where $U_0$ in Eq. \eqref{etaeq} is determined by solving Eqs. \eqref{etaeqa} and \eqref{etaeqb}. Finally, the parameter $\eta_{\rho_B}$ that appears in Eq. \eqref{alphaee} can now be obtained by evaluating the function $\eta(U)$ at $U=U_0$.

For $2 t_B>U_B>-2t_B$ ($-1< \Delta < 1$) one uses instead the parametrization $\Delta = -\cos \gamma$ and obtains similar equations, with the hyperbolic functions replaced by their trigonometric counterparts.

\subsection{$\rho_B=1/3$ state in the constrained bosonic model \eqref{EH}}
\begin{figure}[th]
	\includegraphics[width=0.5\linewidth]{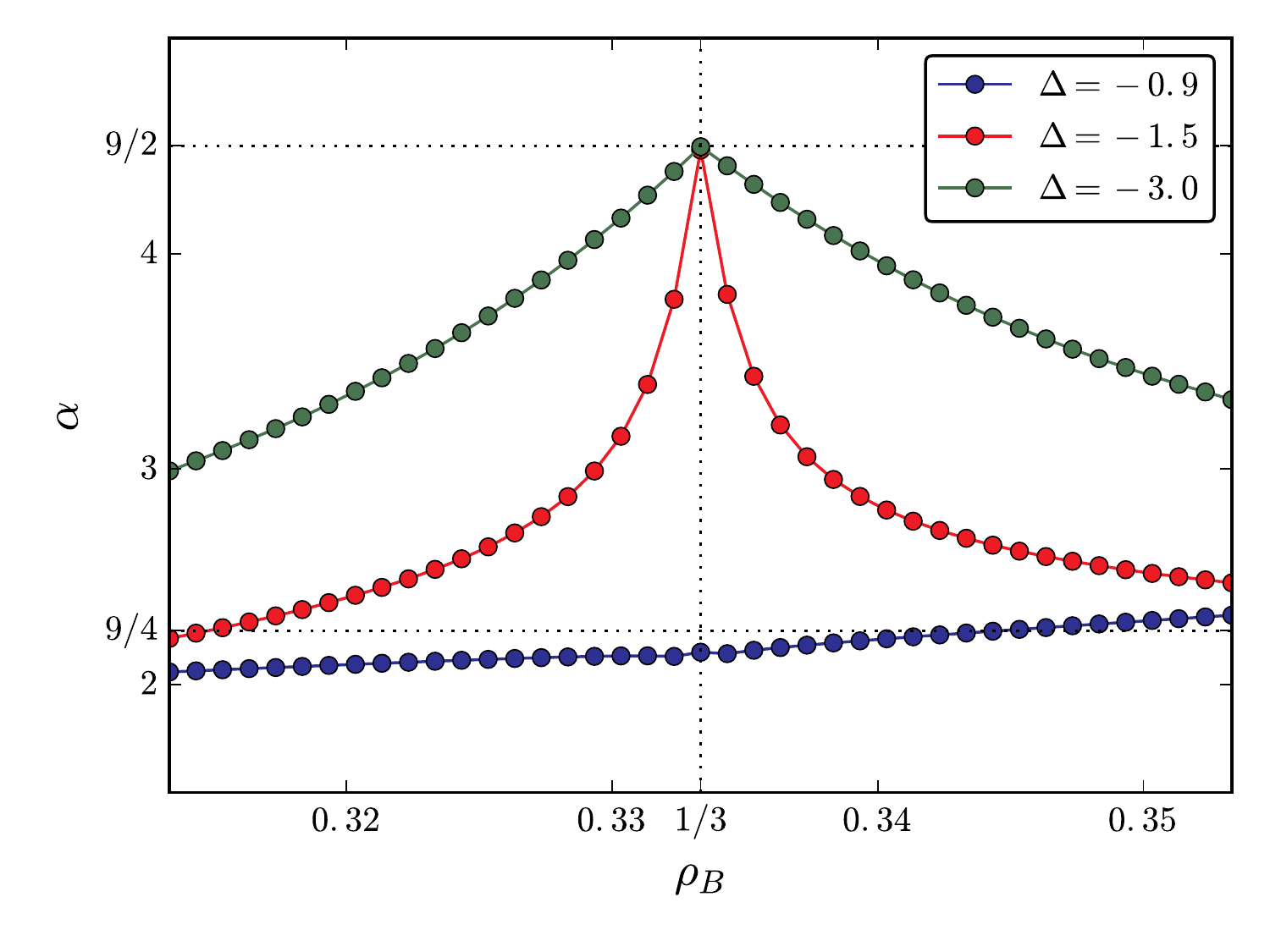}
	\caption{Exponent for the power-law decay of the two-point correlator in the constrained model \eqref{HXXZ} in the vicinity of the commensurate density $\rho_B = 1/3$.}
	\label{fig:alphas_alc}
\end{figure}
After determining $\eta_{\rho_B}$ from Eq.  \eqref{etaeq}, \eqref{etaeqa},\eqref{etaeqb}, one observes that for $U_B > 2t_B$ the exponent $\alpha(\rho_B)$ has a cusp-like peak at $\rho_B = 1/3$, see Fig. \ref{fig:alphas_alc}. The value of $\alpha$ at this point is independent of $U_B/t_B$ and equals to $9/2$. However as $U_B \rightarrow 2 t_B$ from above, the width of the peaks goes to zero. On the other hand, as $U_B \rightarrow 2 t_B$ from below,  $\alpha \to 9/4$. As a result, at the filing $\rho_B=1/3$ the value of the exponent $\alpha$ is a discontinuous function of the ratio $U_B/t_B$. Physically, at the density $\rho_B = 1/3$ the constrained bosonic model is in the $Z_3$ Mott state for $U_B >2t_B$, while it forms a Luttinger liquid for $U_B\le 2 t_B$. The narrowing of the peak corresponds therefore to the closing of the Mott gap as the critical value $U_B=2t_B$ is approached from above. In the Luttinger liquid language, $\alpha = 9/2$ corresponds to a value $K = 1/9$ for the Luttinger parameter. This is exactly the value of $K$, where the commensurate-incommensurate (Mott-$\delta$) transition into a Mott phase of commensurability $3$ takes place \cite{Giamarchibook}.   

\subsection{Mapping to spin-$1/2$ Heisenberg chain in squeezed space and emergent deconfined excitations at $h\gg t$}
When $h=0$, the bare fermions $f_i$ describe deconfined excitations carrying one unit of the $U(1)$ charge. Here we discuss the other extreme, $h \gg t$, where the bare fermions are confined into tightly bound dimers carrying two units of the $U(1)$ charge. The interactions between these dimers, in turn, lead to the formation of a collective Luttinger liquid phase which has fractionalized collective excitations carrying one unit of the $U(1)$ charge. We will demonstrate this now by establishing explicitly a mapping of our model with $h \gg t$ to the spin-$1/2$ Heisenberg model.

When $h / t \to \infty$ we can restrict ourselves to the set of basis states with $\mathbb{Z}_2$ electric field lines ($\sigma^x_{i,i+1}=-1$) of length one, with one fermion at each end. We can label these basis states by the positions $x_1 < x_2 < ... < x_N$ of the fermions, subject to the constraint that 
\begin{equation}
x_{2n} = x_{2n-1}+1 \qquad \text{for} ~ n=1 ... N.
\label{eqLargeHconstrnt}
\end{equation}
The $\mathbb{Z}_2$ field configuration follows from the Gauss law, noting that $\sigma^x_{j-1,j} = 1$ for all $j \leq x_1$ left of the first fermion. An example is illustrated in Fig.~\ref{fig:squeezed_space} (a). 

\begin{figure}[th]
	\includegraphics[width=0.8\linewidth]{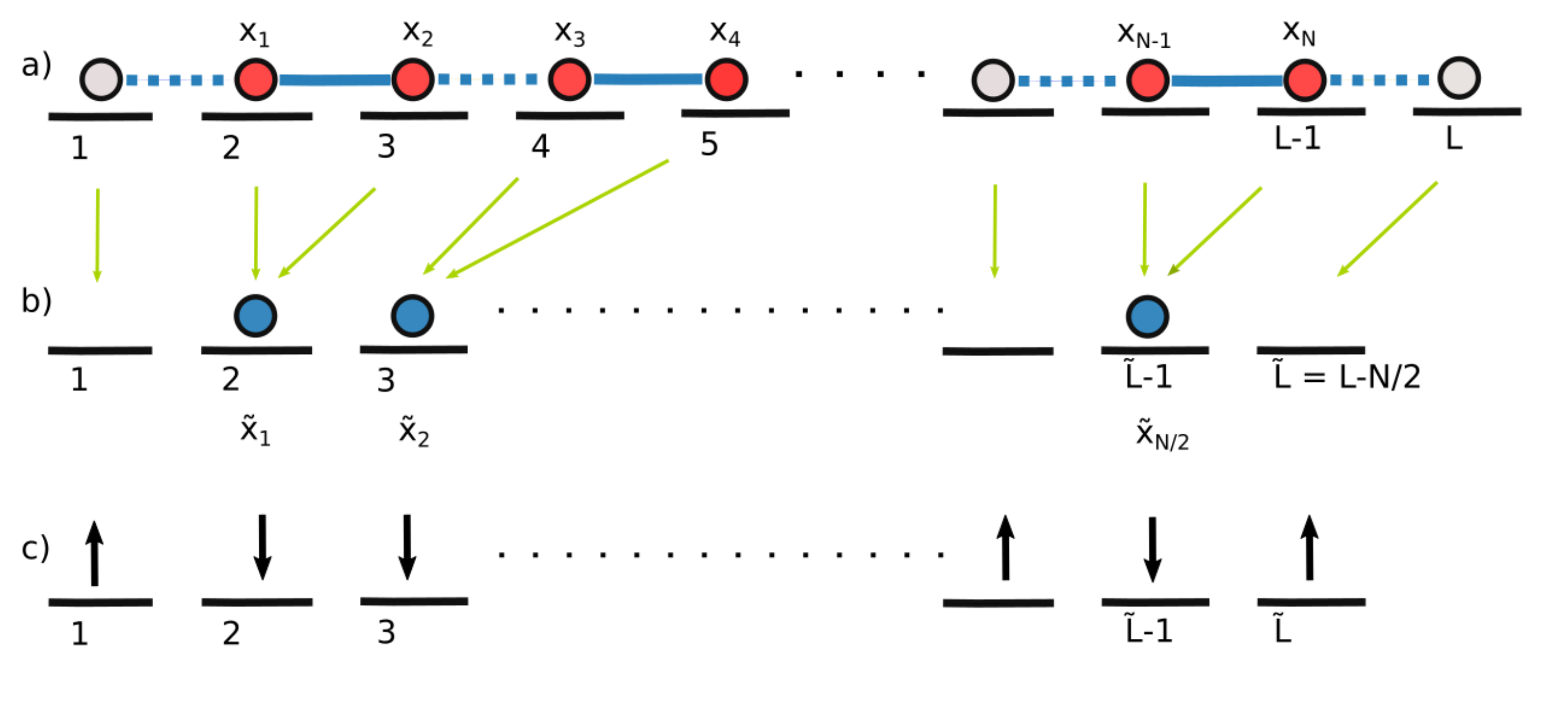}
	\caption{Mapping of compact dimers in $Z_2$ lattice gauge theory to spins in squeezed space.}
	\label{fig:squeezed_space}
\end{figure}

Now we will introduce a new set of labels for these basis states $\ket{x_1,...,x_N}$, by working in a squeezed space (this mapping is motivated by related constructions in doped spin chains \cite{Ogata1990, *Kruis2004a} ): To this end we remove every second fermion from the chain, see Fig.~\ref{fig:squeezed_space} (b):
\begin{equation}
\ket{x_1, ... , x_N} \equiv \ket{\tilde{x}_1, ... , \tilde{x}_{N/2}}, \quad 0 \leq \tilde{x}_j \leq \tilde{L} = L - N / 2. 
\end{equation}
Here $L$ denotes the length of the chain and, without loss of generality, we consider the case when the total fermion number $N$ is even. Using the constraint Eq.~\eqref{eqLargeHconstrnt}, the configuration $x_1, ... , x_N$ can be reconstructed from the squeezed space configuration $\tilde{x}_1 < ...  < \tilde{x}_{N/2}$:
\begin{equation}
x_{2n} = \tilde{x}_n + n, \quad x_{2n-1} = x_{2n} - 1,
\end{equation}
with $n=1 ... N/2$. The reverse is also true, establishing the one-to-one correspondence between the two basis representations.

It is more convenient to work in second quantization, so we introduce hard-core bosons $\tilde{b}_j^{(\dagger)}$ in squeezed space, with $j=1...\tilde{L}$. The basis states thus read
\begin{equation}
\ket{\tilde{x}_1, ... , \tilde{x}_{N/2}} = \prod_{n=1}^{N/2} \tilde{b}^\dagger_{\tilde{x}_n} \ket{0}.
\end{equation}
Performing perturbation theory in $t/h$ up to second order, we arrive at the following effective Hamiltonian in squeezed space:
\begin{equation}
\hat{\mathcal{H}}_{\rm eff} = - t_{\rm B} \sum_{j=1}^{\tilde{L}-1} \left( \tilde{b}^\dagger_{j+1} \tilde{b}_j + {\rm h.c.} \right) + U_{\rm B} \tilde{n}^b_{j+1} \tilde{n}^b_j.
\end{equation}
As in the main text, $t_{\rm B} = t^2 / 2 h$ and $U_{\rm B} = 2 t_{\rm B}$, and we introduced $\tilde{n}^b_j = \tilde{b}^\dagger_j \tilde{b}_j$.

The hard-core bosons $\tilde{b}_j$ can be mapped to spin-$1/2$ operators $\tilde{\mathbf{S}}_j$ in squeezed space in the usual way,
\begin{flalign}
\tilde{S}^z_j &= \left( 1/2 - \tilde{n}^b_j \right),\\
\tilde{S}^-_j & =  \tilde{b}^\dagger_j \\
\tilde{S}^+_j & =  \tilde{b}_j. 
\end{flalign}
Therefore, up to a constant overall energy shift and after changing $\tilde{S}^{x,y}_{2i} \to - \tilde{S}^{x,y}_{2i}$ on even sites, we obtain
\begin{equation}
\hat{\mathcal{H}}_{\rm eff} = U_{\rm B} \sum_{j=1}^{\tilde{L}-1} \tilde{\mathbf{S}}_{j+1} \cdot \tilde{\mathbf{S}}_j.
\end{equation}
Hence asymptotically when $h \gg t$ the problem maps to the $SU(2)$ invariant spin-$1/2$ Heisenberg model in squeezed space, with anti-ferromagnetic coupling $U_{\rm B}$.

It is well-known that the  collective excitations of the spin-$1/2$ antiferromagnetic Heisenberg chain are fractionalized fermionic spinons \cite{Giamarchibook, Fradkin2013}, carrying spin $1/2$. Since the creation of a new dimer of two units of $U(1)$ charge -- by applying $\tilde{b}^\dagger_j$ -- corresponds to a spin-flip with $\Delta \tilde{S}^z =  \pm 1$ in our mapping, it follows that the spinon excitations carrying $\tilde{S}^z= \pm 1/2$ correspond to unit-charged collective excitations in the original $Z_2$ lattice gauge theory.

\subsection{Central charge from entanglement-entropy scaling}
The iDMRG algorithm that we use allows to readily access the entanglement entropy $S$ of a semi-infinite chain and the correlation length $\xi$. For a gapless system both these quantities are in principle infinite, but performing iDMRG at a finite bond dimension $\chi$ effectively introduces a cutoff length which makes them finite. As the maximum bond dimension is increased (cutoff lowered), both quantities grow following the CFT relation
\begin{equation}
S = \frac{c}{6}\log \xi + a,
\end{equation}  
where $c$ is the central charge of the CFT and a is some non-universal constant. Therefore, by looking at how $S$ and $\log \xi$ scale as the bond dimension is increased allows to extract the central charge of the CFT corresponding to our model.
\begin{figure}[th]
	\includegraphics[width=0.8\linewidth]{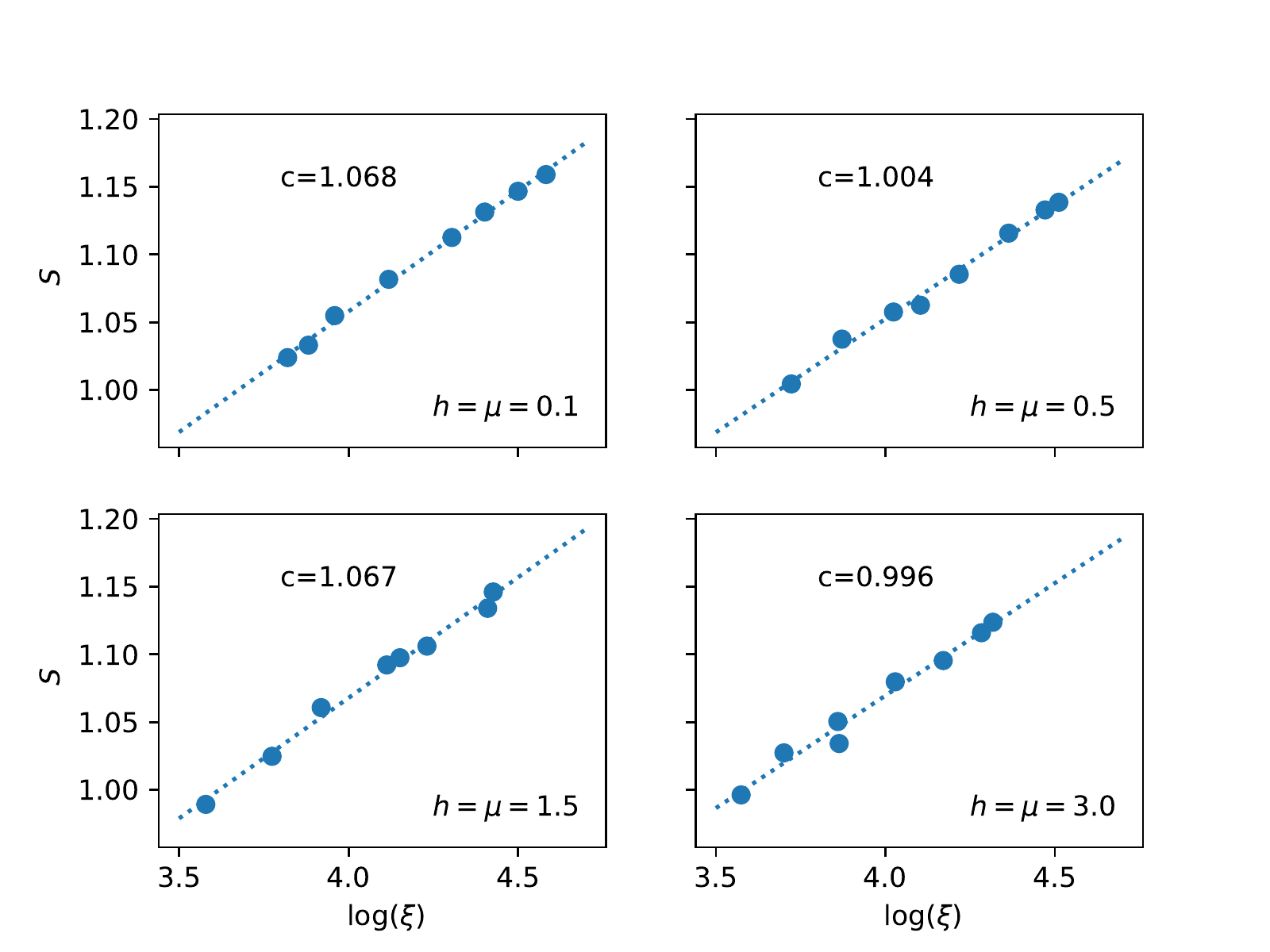}
	\caption{As the bond dimension is increased, the entanglement entropy scales linearly with $\log \xi$. The slope of the line is $c/6$.}
	\label{fig:cc}
\end{figure}
The results shown in Fig. \ref{fig:cc} suggest that our system is a gapless Luttinger liquid with $c=1$ for arbitrary values of the string tension $h$.

\subsection{Friedel Oscillations}
As explained in the main text, an important feature of the system is the doubling of the period of Friedel oscillations in a chain with open boundary conditions. Interestingly, such doubling takes place as soon as the coupling $h$ to the gauge fields is turned on, and the profile of the oscillations does not undergo significant variations as $h$ is increased (See Fig. \ref{fig:friedel_all}).
\begin{figure}[th]
	\includegraphics[width=0.8\linewidth]{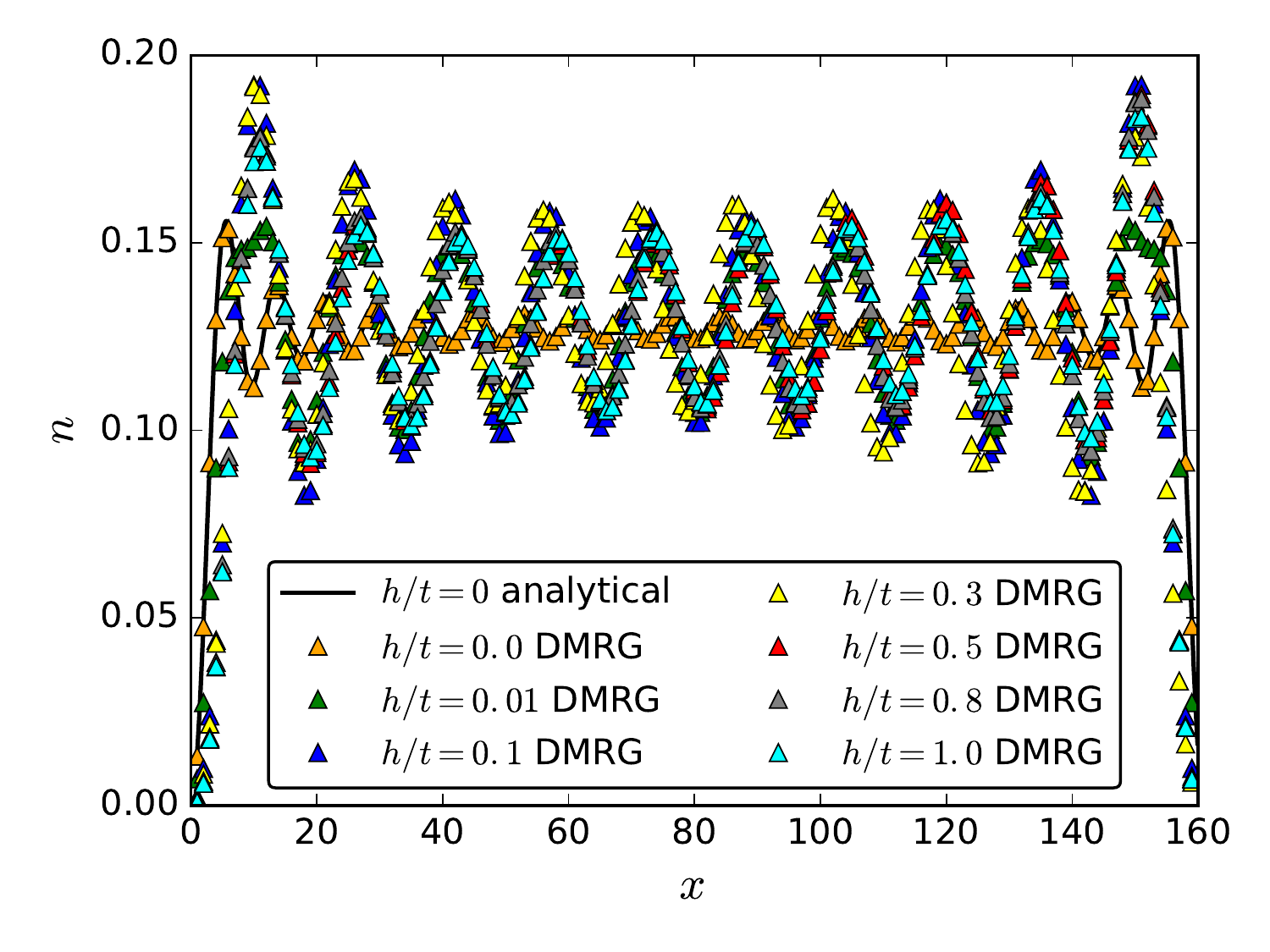}
	\caption{Friedel oscillations in a chain of length $L=160$ at filling $1/8$.}
	\label{fig:friedel_all}
\end{figure}
This behavior is to be compared with the one of the Luttinger parameter $K$, which is a continuous function of $h$ at any given filling. For instance, at small $h=0.01$ one has $K \approx 1$ (See Fig. \hyperref[fig:combined]{3c}), which is the value expected for free fermions, in agreement with our field theory analysis. For such value of $h$, however, the doubling already occurs and signals the confinement of \textit{lattice} fermions.

\subsection{Average electric string length from two-particle Hamiltonian}\label{str_length}
In our problem two $Z_2$ charges are always connected by an electric string. Due to the competition between the energy cost $\propto h$ of the string and the kinetic term, which tends to delocalize fermions, the string has an average length $\langle l \rangle$ which is a decreasing function of $h/t$. Since the center-of-mass motion of the two particles can be separated out, $\langle l \rangle$ can be computed from the ground state of the following single-particle Hamiltonian
\begin{equation} \label{rel}
H = -2 t \sum_{l=1}^{\infty}(f^{\dagger}_{l+1}f_{\,l}^{\vphantom \dagger}+\text{h.c})+2 h \sum_{l=1}^{\infty} l \, f^{\dagger}_{l}f_{\,l}^{\vphantom \dagger}.
\end{equation}  
We find that $\langle l \rangle$ is rather small (of order of units of the lattice spacing) even for moderate values of $h/t$. For large values of $h/t$, it goes to one as a power law, see Fig. \ref{fig:st_length}. The dimers can be thought as sufficiently separated molecules provided their size is much smaller than the average interparticle distance, i.e.,  $\langle l \rangle \ll \rho_F^{-1}$.
\begin{figure}[th]
	\includegraphics[width=0.5\linewidth]{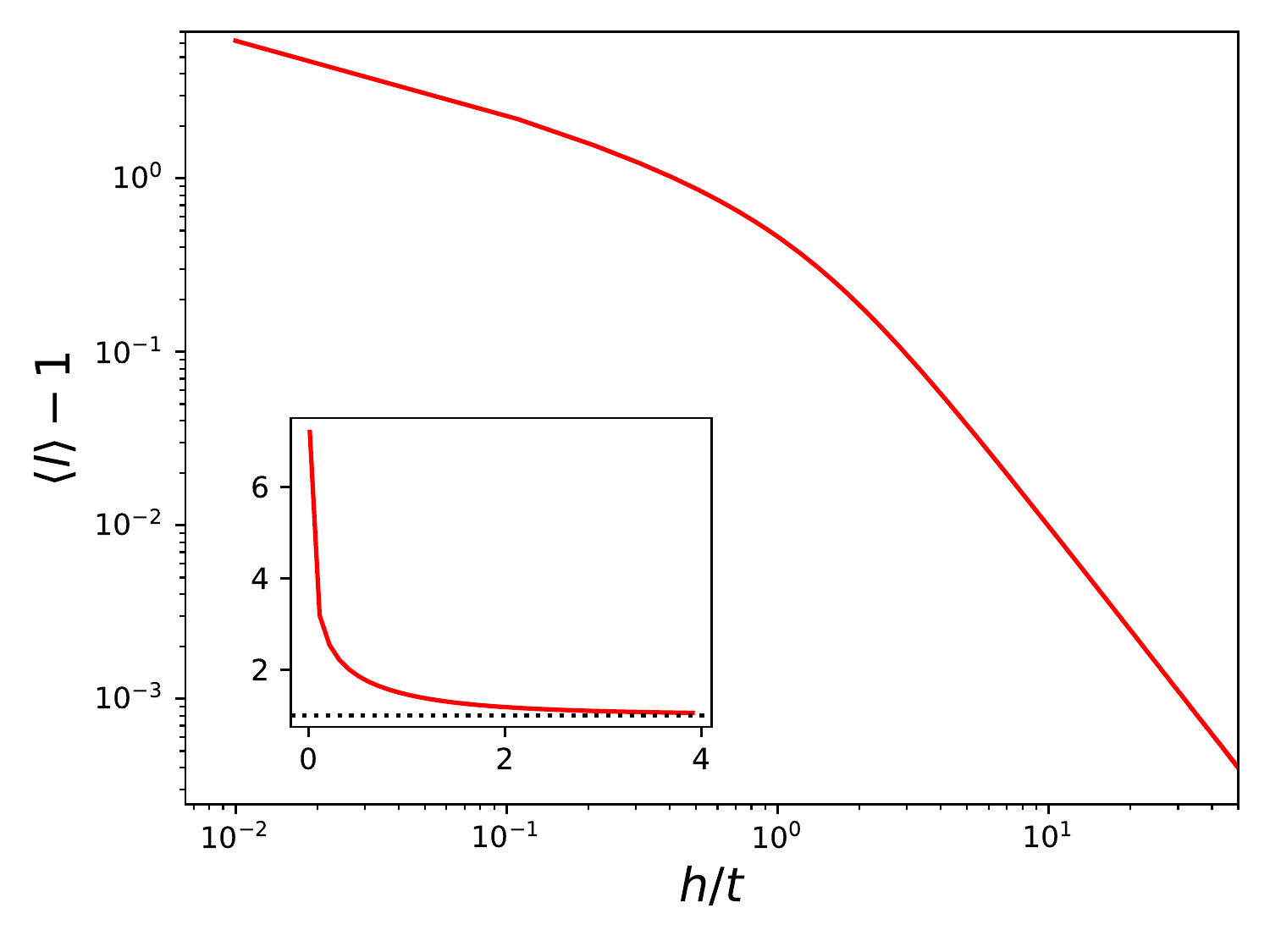}
	\caption{The average string length $\langle l \rangle$ as a function of $h/t$ extracted from the Hamiltonian \eqref{rel}.}
	\label{fig:st_length}
\end{figure}


\end{widetext}

\end{document}